\documentclass[aps,prb,twocolumn,eqsecnum]{revtex4}
\usepackage{amsmath}
\usepackage{amssymb}
\usepackage{graphicx}
\usepackage{xcolor}
\usepackage{dcolumn}
\usepackage{bm}

\def\be{\begin{equation}}
\def\ee{\end{equation}}
\def\bea{\begin{eqnarray}}
\def\eea{\end{eqnarray}}

\begin{document}
\title{Non-equilibrium field theory with a temperature gradient: Thermal current in a nanowire}
\author{Yuan Gao and K.A. Muttalib}
\affiliation{Department of Physics, University of Florida, P.O. Box 118440, Gainesville, FL 32611-8440}

\today

\begin{abstract}
A perturbative framework is developed within the standard non-equilibrium field theory techniques to incorporate a temperature gradient across a thermoelectric device. The framework uses a temperature-dependent pseudo-Hamiltonian generated from the exact density matrix in the presence of non-uniform temperatures. We develop a perturbation theory  for small temperature gradients in long wires and obtain the non-linear thermal conductance as a function of temperature difference. The framework should be adaptable to more general cases of temperature inhomogeneities in either fermionic or bosonic systems.

\end{abstract}
\maketitle

\newpage


\section{Introduction}

Thermoelectric current is generated by converting heat to electrical energy \cite{book}. A typical thermoelectric device is a good electrical conductor with its two ends kept at two different fixed temperatures, generating a difference in chemical potential between the two ends  \cite{review1, review2, review3}. However, the device also needs to be a poor thermal conductor in order to have a large efficiency. It has been proposed \cite{mm21} that improved efficiency and power output in thermoelectric devices require two independent considerations: Taking advantage of an interplay between the material and the thermodynamic parameters available to increase the charge transport in the \textit{non-linear} regime \cite{hmn13, mh15}, and simultaneously decreasing the heat current by using surface disordered nanowires \cite{boukai08, li03, hochbaum08, chen08, lim12, heron09, blanc13, blanc14}. Because of the temperature difference between its two ends, there exists a temperature gradient across the nanowire. In the linear response regime, the temperature difference is assumed to be small and one can treat the entire device at one fixed average temperature. In this case one can use the standard Non-Equilibrium Field Theory (NEFT) techniques \cite{rammer, dhar12}, valid for a thermally homogeneous system, to evaluate both the electrical and the thermal current in the device. In the non-linear regime mentioned above, the temperature difference between the ends can be large and the temperature of the wire is different at different distances from the two ends. The standard NEFT techniques breakdown in such cases; studying charge or heat currents requires a generalization of the currently available NEFT methods to incorporate thermal inhomogeneity.  

In this paper we generalize the existing NEFT techniques in the presence of a temperature gradient and evaluate the heat current across a thermoelectric device. For simplicity we consider a uniform wire of length $L=Na$, $a$ being the lattice spacing and the number of sites $N \gg 1$. One end of the wire is kept at a fixed hot temperature $T_H$ and the other end at a colder temperature $T_C$. We use a perturbation theory valid for small temperature \textit{gradients}. The framework divides the system into slices of subsystems at different temperatures connected together and allows us to describe the non-equilibrium Green's functions in terms of the temperature at the mid-point and additional temperature-dependent perturbation terms proportional to the thermal gradient parameter $\gamma /L$, with $\gamma \equiv \ln T_H/T_C$. The phonon Green's functions can be explicitly written down in terms of the same parameter and the temperature of the midpoint of the conductor. We show that for $\gamma \ll 1$ and $N \gg 1$, a systematic and consistent perturbation theory can be developed that allows us to study the effects of the temperature gradient on the thermal current in the absence of any other perturbation. Further perturbations can then be added in this inhomogeneous thermal system.

While we consider a long nanowire as the simplest inhomogeneous thermal system for illustrative purposes, it should be possible to generalize the framework to include arbitrary thermal inhomogeneities. The method involves re-interpreting the density matrix of a given Hamiltonian with different subsystems at different temperatures in terms of an effective density matrix of a new temperature-dependent pseudo-Hamiltonian at one arbitrary but fixed temperature.  In the following sections we first introduce the framework for general inhomogeneous systems.  We exploit the fact that Wick's theorem can be applied if we incorporate  the thermal inhomogeneity in the density matrix via a pseudo-interaction term added to our Hamiltonian. We consider the case of phonons in detail which is relevant for thermoelectric devices. While the inhomogeneity generates several additional (pseudo) interaction terms, we illustrate our method by choosing one novel term and evaluating the corresponding effects on the non-equilibrium Green's functions as well as the thermal current  within our perturbative framework. We evaluate the frequency dependent thermal current as well as the non-linear thermal conductivity for finite temperature differences to leading order in $\gamma$. These effects should be observable in thermal transport measurements in nanowires. 

The paper is organized as follows. We introduce the general framework for thermally inhomogeneous systems in Section II and obtain the correction to the phonon Greens function to first order in the parameter $\gamma$  due to a thermal gradient across a nanowire in Section III. The resulting thermal current and the corresponding thermal conductance are obtained in Section IV. Section V contains summary and discussion.


\section{Framework for thermally inhomogeneous  systems}  

\begin{figure}
\includegraphics[angle=0,width=0.4\textwidth]{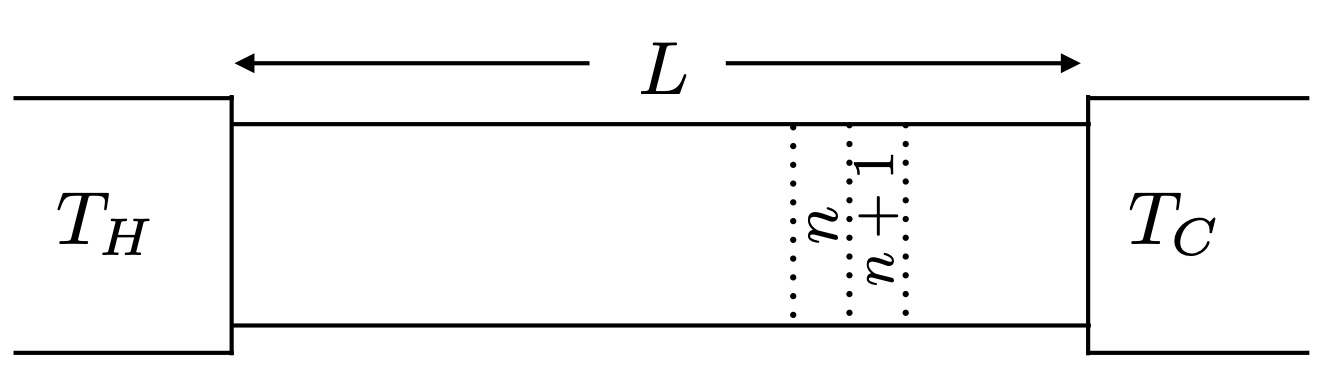}
\caption{ A wire of length $L$ attached to two leads, hot and cold, with temperatures $T_H$ and $T_C$, respectively. The temperature inside the wire has a gradient along the length. We imagine dividing the wire  into thin slices, the $n$th slice being at inverse temperature $\beta_n$. }
\label{Fig-TempProfile}
\end{figure}

For simplicity, we consider a wire of length $L$ as shown in Figure \ref{Fig-TempProfile} with two ends connected to perfect leads fixed at two different temperatures $T_H$ and $T_C$;  the subscripts refer to hot and cold, respectively. 
The temperature difference creates a stationary thermal gradient inside the wire. We now imagine dividing the wire into $N$ number of thin slices, the $n$th slice being at temperature $T_n$. The Hamiltonian of the system can be written as a collection of $N$ subsystems
\bea
H=\sum_n [ H_n^0+H_n^I +H^I_{n,n+1} ]
\label{H}
\eea
where $H_n^0$ and $H_n^I $ are the free (quadratic) and the interaction parts of the $n$th slice, respectively, and $H_{n,n+1}$ is the interaction between connected slices. The corresponding density matrix is given by
\bea
\rho(t_0)=\frac{\exp[-\sum_n\beta_n(H_n^0 + H_n^I + H_{n,n+1}^I )]}{Tr (\exp[-\sum_n\beta_n(H_n^0+ H_n^I + H_{n,n+1}^I )])} .
\label{rho-H}
\eea
The Greens function can be written as 
\bea
G(1,2)=-i Tr [\rho(t_0){\cal T} \psi_H(1) \psi^{\dag}_H(2)]
\eea
where $\cal{T}$ is the time-ordering operator.

The fact that each slice has a different temperature makes the evaluation of the Green's functions difficult because a compact time-ordering is not available and the Wick's theorem cannot be applied directly. 
The fact that the density matrix in (\ref{rho-H}) doesn’t commute with the Hamiltonian (\ref{H}) suggests that a thermally homogeneous Hamiltonian is not a good description of the time evolution of a thermally inhomogeneous system. We take this into account by rewriting the numerator of the density matrix as
\bea
&&\sum_n\beta_n[H_n^0 + H_n^I + H_{n,n+1}^I ] = \beta_0 H_{p} \cr
H_{p} &\equiv &  \sum_n \left[ H_n^0 + \frac{\beta_n-\beta_0}{\beta_0}H_n^0 + \frac{\beta_n}{\beta_0} (H_n^I+H_{n,n+1}^I)\right] .\cr
&&
\label{Hp}
\eea
We now observe that the density matrix of  $H$  defined in (\ref{H}) with different inverse  temperatures $\beta_n$ at different slices $n$ is the same as the density matrix of a pseudo-interaction Hamiltonian $H_{p}$ in (\ref{Hp}) at an arbitrary but fixed inverse temperature $\beta_0$.  We  choose to work with the temperature-dependent pseudo-Hamiltonian $H_{p}$ as our starting point, which means we do all the traces on eigenstates of $H_{p}$; the perturbation theory for this pseudo-Hamiltonian can then be worked out  within the standard NEFT framework, albeit with additional non-standard temperature-dependent interaction terms. In the following section, we will illustrate the method with a simple but important example, taken from thermoelectricity, where a clear small expansion parameter can be identified and the perturbation theory becomes well-defined in the large $N$ limit.

We note that a different series expansion for a thermally inhomogeneous system was proposed in [\onlinecite{langmann17}] and Wick’s theorem seems to work for an interacting system within the Equilibrium Green’s Function (EGF) framework. There are also exact solutions available for some expectation values in one-dimensional models that can be mapped on to conformal field theories \cite{moosavi1, moosavi2}. Our framework should be more generally applicable for both EGF and NEFT.


\section{Nanowire with a temperature gradient} 

As a simple illustrative example, we consider phonon propagation across a nanowire. 
We assume that the free phonon dispersion relation is given by 
\be
\omega_k^2=\frac{K}{\rho a^3}\sum_{i=x,y,z} \cos k_i a 
\ee
where $\rho$ is the mass density, $K$ is the spring constant.  We model the temperature profile in the wire in the continuum limit given by  
\be
\beta_n  \rightarrow  \beta(x)=\beta_0e^{\gamma x/L},
\ee
 with
\bea
T_0 = \sqrt{T_HT_C}; \;\;\; e^{\gamma}=\frac{T_H}{T_C}.
\label{temperature-profile}
\eea
In the small $\gamma$ limit, to linear order in $\gamma$, this corresponds to a uniform temperature gradient across the length of the wire, which we take to be the $x$ direction. 
 We note that in a clean strictly one-dimensional wire, the temperature profile remains flat in the bulk, any changes occurring only at the edges \cite{dhar03, rieder67}.
However, measurements on thin nanowires show a temperature profile similar to above \cite{hoffmann09}.
For these nanowires, the process to achieve a temperature gradient due to the phonon anharmonicity, phonon-impurity scattering and/or electron-phonon interactions and for the particles to reach non-equilibrium steady states after the attachment of two leads with different temperature could be complicated. However, after the temperature gradient is formed, the perturbation brought by the temperature inhomogeneity contributes to the phonon Green’s functions in first order, compared to  the second order contribution from the other scattering sources mentioned above. We therefore start from a free-phonon model and treat the temperature inhomogeneity as the dominant perturbation term. The second order effects can be included in the future, adding to the novel contribution from the temperature inhomogeneity alone. In any case, an exact temperature profile is not important to illustrate the applicability of our framework; we therefore choose the simplest case, leaving a more accurate account of the temperature profile for future work. 
Figure  \ref{Fig-TempProfile} shows the system being considered, where $\Delta T=T_H -T_C $ is not necessarily small 
(for example $T_C=300 K$, $\Delta T = 30 K$), allowing us to develop a perturbative framework with $\gamma  \sim \Delta T/T_C\ll 1$ ($\gamma =0.1$ in the above example) being the expansion parameter.

In the continuum limit the quadratic part is
\be
H^0 =\frac{1}{2} \int \vec{dr} [ \dot{u}(\vec{r})  \dot{u}(\vec{r}) 
+ \frac{V}{2} u(\vec{r})\{u(\vec{r}-\vec{a})+ u(\vec{r}+\vec{a})\}]
\ee
where the overdot represents a time derivative. Here $V=K/\rho a^3$, $u=\sqrt{\rho} U$, $U$ being the displacement of the atoms from the equilibrium positions, and $\vec{a}$ are the lattice vectors.   In the simplest case, the perturbative pseudo-interaction term generated by the temperature gradient along $x$ is 
\bea
H^{pert} &=& \frac{\gamma}{2L}\int \vec{dr} \;x \; [\dot{u}(\vec{r}) \dot{u}(\vec{r}) \cr
&+& \frac{V}{2}  \; u(\vec{r})\{u(\vec{r}-\vec{a})+ u(\vec{r}+\vec{a})\}].
\label{HI}
\eea
 Note the explicit  $x$-dependent terms reflecting the thermal inhomogeneity of the system, where we chose the midpoint of the wire as the origin $x=0$.  The most interesting part is the term with $\vec{a}$ along $x$, and we will illustrate our method with a focus on this novel term. Note also that the interaction term is proportional to $\gamma/L$, not just $\gamma$.
 
 We note that the linearized temperature profile $\beta(x) = \beta_0[1+\gamma x/L]$ looks similar to the position-dependent field that particles couple to in the Luttinger formalism \cite{luttinger64} with $\beta(r) =\beta_0[1+ \psi(r)]$ as discussed in [\onlinecite{cooper97}]. Our framework, derived directly from the exact density matrix, clarifies the physical meaning of the fictitious ‘mechanical’ field $\psi$, and with the explicit form of the field coming out of the temperature profile, one can conveniently go beyond the linear-response regime.
  
The phonon Green's function can be written as 
\bea
D(1,2)=-i\; Tr [\rho(t_0){\cal T} u_{H_p}(1) u_{H_p}(2)] .
\eea
Here we have used the phonon displacement fields instead of the raising and lowering operators since the interaction (\ref{HI}) has an explicit $x$ dependence. In addition, it is useful to define a correlation function 
\bea
C^{0} (1,2) = Tr [\rho(t_0){\cal T} u_{H^0}(1) \dot{u}_{H^0}(2)] .
\eea
The zeroth order phonon Green's functions $D^{0}(\vec{p},t-t')$ are well-known and the correlation functions $C^{0}(\vec{p},t-t')$ can be easily calculated. The first order contribution to the phonon Green's function induced by the temperature gradient  via the pseudo-interaction term $H^I$, with $\vec{a}$ along $x$, is then given  by 
\bea
&& \delta^{(1)}D(x, x', \vec{k }; t, t')= (-i) \frac{\gamma}{2L}  \int dt_1\;dq\;\delta(q) \cr 
&&  \{e^{iqx}  [(i x+\partial_q) (\omega_k^2 D^0(\vec{k}+\vec{q},t-t_1)D^0(\vec{k}, t_1-t') \cr
&&\;\;\;\;\;\;- C^0(\vec{k}+\vec{q},t-t_1)C^0(\vec{k}, t_1-t')) ]\cr
&& + e^{iqx'} [(i x'+\partial_q) (\omega_k^2 D^0(\vec{k},t-t_1)D^0(\vec{k}-\vec{q}, t_1-t') \cr
&&\;\;\;\;\;\;- C^0(\vec{k},t-t_1)C^0(\vec{k}-\vec{q}, t_1-t')) ]\cr
&& +a V \sin k_xa \;D^0(\vec{k},t-t_1)D^0(\vec{k}-\vec{q}, t_1-t')]\}.
\label{delta1D-FT}
\eea
After the $q$-integral, and going to the frequency space, one can identify the retarded self-energy up to linear order in $\gamma$ by comparing with the Dyson expansion ($D^R=D^{0R}\Sigma^R D^R$)  as
\bea
\Sigma^R(x,x'; \vec{k}, \omega)=\frac{\gamma}{L} \left[ \frac{x+x'}{2} \{\omega_k^2-(\omega+i\delta)^2\} \right. \cr
\left. -ia\frac{V}{2} \sin k_xa \right] .
\label{Sigma}
\eea
The corresponding retarded phonon Green's function takes the form
\bea
D^R &=&\frac{Z}{(\omega+i\delta)^2-\omega_k^2+i\frac{\gamma a}{L} Z \frac{V}{2}\sin k_xa}, \cr
Z &\equiv &  \frac{1}{1+\frac{\gamma}{L}\frac{x+x'}{2}} .
\label{Z}
\eea
Note that the imaginary part in the denominator of $D^R$ is proportional to $\gamma a/L$, where $\gamma \ll 1$ is already a small parameter. In addition, we assume $L \gg a$  so that the combination $\gamma a/L=\gamma/N \ll \gamma$ can be considered negligible in the large $N$ limit. While the specific form of the imaginary part in (\ref{Z})  is model dependent, the prefactor $\gamma a/L$ (in the leading order perturbation theory) ensures that the imaginary part remains negligible. Thus the final result for the Green's function, to first order in $\gamma$ and in the large $N$ limit, becomes
\bea
D^{R/A} =\frac{Z}{(\omega\pm i\delta)^2-\omega_k^2} .
\eea
It is interesting that the numerator $Z$ defined in (\ref{Z}) mimmicks the quasi-particle weight factor in the standard equilibrium fermionic many-body quantum field-theory. Note that at this level of approximation, the consistency relations between various self energies (retarded, advanced, greater and lesser) remain valid: 
$
\Sigma^R=(\Sigma^A)^*;\;\;\; \Sigma^R-\Sigma^A=\Sigma^> - \Sigma^< .
$
We should emphasize that the consistency relations may not be automatically satisfied at each order within our perturbative  formulation with the small parameter $\gamma \ll 1$ alone; the additional small parameter $a/L \ll 1$ (equivalently $N\gg 1$) in our current example is critical.  On the other hand, these relations should be valid if an exact summation to all orders is available. The first order corrections to the lesser and greater Green's functions then have the same common $Z$-factor:
\bea
&&D^{\gtrless}  =  Z D^{0\gtrless}  =  -i Z \frac{\pi}{\omega_k }\times \cr
&&   [ (N_{k}+1) \delta (\omega\mp \omega_{k}) + N_{k} \delta(\omega\pm \omega_{k})] 
\label{gtrless}
\eea
where $N_k$ are the Bose distribution functions.

In particular, the phonon number density given by 
\bea
n(\vec{r}) &=& i\; G^<(\vec{r}'\to \vec{r}, t'\to t) \cr
&=& \int \frac{d\vec{k}}{(2\pi)^3} \frac{1}{1+\gamma x/L} N_k
\label{nofr}
\eea
where $G$ refers to the phonon Greens function involving the raising and lowering operators instead of the displacement fields. (\ref{nofr}) has the simple interpretation that the hotter parts ($x < 0$) have more phonons than the cooler parts ($x >0$).


\section{Thermal conductance} 
\begin{figure}
\includegraphics[angle=0,width=0.4\textwidth]{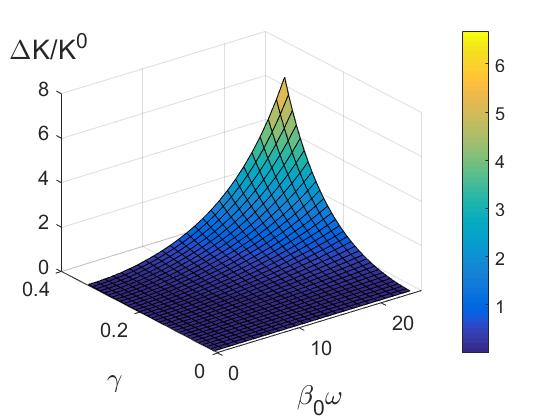}
\caption{The ratio  $\Delta K/K^0 = [K^{\gamma}(\omega)-K^0(\omega)] / K^0(\omega)$ obtained from (\ref{kappa}), plotted as functions of the  parameters $\gamma $ and  $\beta_0\omega$.}
\label{Fig-J}
\end{figure}

\begin{figure}
\includegraphics[angle=0,width=0.4\textwidth]{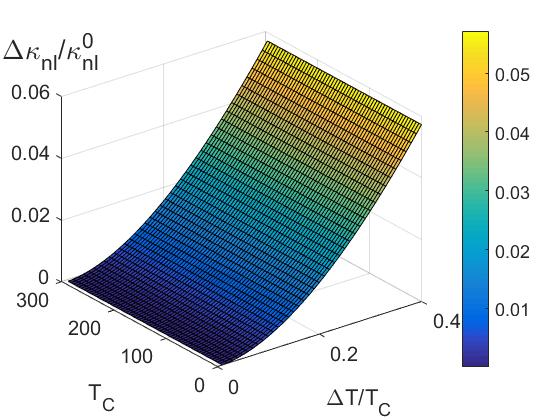}
\caption{The ratio $ \Delta \kappa_{nl}/\kappa^0_{nl} = [\kappa^{\gamma}_{nl} - \kappa^0_{nl}]/\kappa^0_{nl} $ obtained from (\ref{kappa}), plotted as functions of $\Delta T/T_C$ and $T_C$.  }
\label{Fig-kappa}
\end{figure}

The usual definition of thermal conductivity assumes a linear response regime with $\Delta T = (T_H - T_C) \to 0$. In a thermoelectric device we want $\Delta T$ to be large. The general frequency-dependent steady state thermal current across such a wire is given by \cite{LS07, Mingo06}
\bea
&&J(\omega) =\frac{\omega}{4} \int d\vec{r}_1d\vec{r}_2\{ [ D_L^{(c) >}(\vec{r}_1,\vec{r}_2; \omega) \Sigma^<_L(\vec{r}_2,\vec{r}_1; \omega) \cr
&& - D_L^{(c) <}(\vec{r}_1,\vec{r}_2; \omega) \Sigma^{>}_L(\vec{r}_2,\vec{r}_1; \omega)]  - [L\to R]\} .
\eea 
Here the superscript (c) on the phonon Green's function $D$ refers to the central region (the wire), with subscripts L and R referring to the ends connecting the left and right leads, respectively.  The subscripts L and R on the self energy $\Sigma$ refer to the left and right leads, respectively,  and $\vec{r}_1, \vec{r}_2$ lie in the lead-central interface. For simplicity, we consider a quasi-one-dimensional wire with independent channels in the lead and compute the contribution from one channel. This simplifies the calculation by allowing us to ignore the spatial integrals. (A proper evaluation including all channels would change the pre-factor to reflect the fraction of the incoming waves being transmitted/reflected across the two barriers that connect the leads with the wire but should be generally independent of $\gamma$.) 
The Greater/ Lesser Green’s functions in the central region are available from the previous section, equation (\ref{gtrless}), and the lead self-energy is defined to be proportional to lead’s free Green’s function, i.e. 
\bea
&&\Sigma^{\gtrless}_{L/R} \propto  D^{0\gtrless}_{L/R}  =  -i \frac{\pi}{\omega_k } \times \cr
&& [ (N^{L/R}_{k}+1) \delta (\omega\mp \omega_{k}) + N^{L/R}_{k} \delta(\omega\pm \omega_{k})]. 
\eea
Now writing the real space Green’s functions in terms of their Fourier components and doing the $k$-integral, the frequency dependent current can be written as
\bea
J^{\gamma} (\omega) &\propto &\frac{n^2(\omega)}{\omega} \left[ \frac{1}{1-\frac{\gamma}{2}} N^L(\omega) 
-  \frac{1}{1+\frac{\gamma}{2}} N^R(\omega)  \right.  \cr
&-& \left. \frac{\gamma}{1-\frac{\gamma^2}{4}} N^0(\omega) \right] \Theta(\omega) + (\omega \to -\omega)
\label{J}
 \eea
where $n(\omega)$ is the phonon density of states, $N(\omega)$ is the Bose distribution function and $\Theta(\omega)$ is the step function; the superscript 0 refers to the mid-point inverse temperature $\beta_0$, while L and R refer to the temperatures at the left and right leads. 
The midpoint temperature $\beta_0$ appears in $J^{\gamma}$ only in the combination $\beta_0\omega$ via the Bose distributions. The proportionality factor involves the coupling of the leads to the wire, which will depend on the system. To compare with experiments, a non-linear thermal conductance has been proposed in [\onlinecite{mm21}], given by 
\be
\kappa_{nl}=\int K(\omega) d\omega; \;\;\; K(\omega)\equiv \frac{J(\omega)}{\Delta T}.
\label{kappa}
\ee
In the following, we choose a constant phonon density of states to obtain the thermal current in (\ref{J}) from acoustic phonons. Figure \ref{Fig-J} shows the ratio  
\be
\frac{\Delta K}{K^0} = \frac{ [K^{\gamma}(\omega)-K^0(\omega)]}{K^0(\omega)}, \;\;\; K^0=\lim_{\gamma\to 0}K^{\gamma} .
\ee
Figure \ref{Fig-kappa} shows the ratio
\be
\frac{\Delta \kappa }{\kappa^0} = \frac{[\kappa^{\gamma}_{nl} - \kappa^0_{nl}]}{\kappa^0_{nl}} .
\ee
 Note that in this case, contributions to $\kappa_{nl}$ derives primarily from $\omega \ll 1/\beta_0$ region of the current in Figure \ref{Fig-J}. On the other hand, contributions from optical phonons with frequency $\omega \sim 20/ \beta_0$, according to Figure \ref{Fig-J},  would result in much larger changes in $\kappa_{nl}$. 
Note that thermal conductivity could be defined in a similar fashion, with an added length dependence. However, since we evaluate the ratio $K/K^0$, the results will be valid for both conductance and conductivity. These predictions can be verified experimentally.

\section{Summary and discussion:} We have developed a perturbative framework within the standard non-equilibrium quantum field theory to study thermal current in a nanowire connected  to hot and cold leads with arbitrary but fixed temperatures $T_H$ and $T_C$, respectively. The framework allows us for the first time to consider  the case $\Delta T = (T_H -T_C)$ not necessarily small, which results in a finite temperature gradient across the wire and conventional field theory with a fixed temperature cannot be applied directly. The framework relies on starting from a temperature-dependent pseudo-Hamiltonian equivalent to an exact density matrix. While we choose to work with a uniform gradient of temperature in a long wire to show the need for and the effectiveness of our method, the framework should be adaptable to more complicated systems, including electrons, with different types of inhomogeneities in temperatures. Of course the method would be practically useful if a small parameter can be identified to develop a perturbation theory, or if the pseudo-Hamiltonian is exactly soluble.

Based on the new framework, we evaluate the frequency dependent thermal current $J^{\gamma}(\omega)$ and the non-linear thermal conductance $\kappa^{\gamma}_{nl}$ in a wire due to the temperature difference $\Delta T$, to leading order in $\gamma$. In the absence of a theoretical framework to incorporate temperature inhomogeneities, experimental works so far have focused primarily on the linear response regime, with $\Delta T \to 0$; we hope that our results would encourage experiments in the finite $\Delta T$ regime relevant for thermoelectric devices. If experiments involve larger $\gamma$, higher order in perturbation or a self-consistent calculation might be needed.

In real systems the wire would have other sources of scattering that might complicate the interpretation of the results. Although it is known that effects of  bulk disorder on thermal conductivity is negligible in the linear response regime, the situation is different in the presence of surface disorder, which generates randomly positioned localized phonons \cite{ ma17, mm18, mm19, am19}. It would therefore be important to include the surface disorder effects in this framework to study the effects of any coupling between the phonons generated by surface disorder and those arising from temperature inhomogeneities. In addition, the framework can be easily adapted to investigate the effects of finite $\Delta T$ on charge transport, needed to study the efficiency of thermoelectric devices.

\textit{Acknowledgements:} We are grateful to Yuxuan Wang for useful discussions and bringing our attention to References [\onlinecite{moosavi1, moosavi2}]. YG Acknowledges partial support from Dr. Emmanuel and Dora G. Partheniades Physics Award for green energy technology.
 

\end{document}